\documentclass[utf8]{FrontiersinVancouver} 
\usepackage{url,hyperref,lineno,microtype,subcaption}
\usepackage[onehalfspacing]{setspace}
\usepackage{lipsum}
\usepackage{multirow}
\usepackage {xcolor}
\usepackage{amsmath}
\usepackage{lineno,hyperref}
\usepackage{multicol}
\usepackage{makecell}

\def\keyFont{\fontsize{8}{11}\helveticabold }
\def\Authors{H.~Lee\,$^{1,2}$,B.J. Park\,$^{1,2}$,J.J. Choi\,$^{2,3}$, O. Gileva\,$^{2}$, C. Ha\,$^{4}$, A. Iltis\,$^{5}$, E.J. Jeon\,$^{2,1}$, D.Y. Kim\,$^{2}$, K.W. Kim\,$^{2}$, S.H. Kim\,$^{2}$, S.K. Kim\,$^{3}$, Y.D. Kim\,$^{2,1}$, Y.J. Ko\,$^{2}$, C.H. Lee\,$^{2}$, H.S. Lee\,$^{2,1}$, I.S. Lee\,$^{2,*}$, M.H. Lee\,$^{2,1}$,  S.J. Ra\,$^{2}$, J.K. Son\,$^{2}$, K.A.~Shin\,$^{2}$}
\def\firstAuthorLast{H.~Lee {et~al.}}
\def\Address{
  $^{1}$IBS School, University of Science and Technology (UST), Daejeon 34113, Republic of Korea\\
  $^{2}$ Center for Underground Physics, Institute for Basic Science (IBS), Daejeon 34126, Republic of Korea\\
  $^{3}$ Department of Physics and Astronomy, Seoul National University, Seoul 08826, Republic of Korea\\
  $^{4}$ Department of Physics, Chung-Ang University, Seoul 06973, Republic of Korea \\
  $^{5}$ Damavan Imaging, Troyes 10430, France
}
\def\corrAuthor{I.S.~Lee}

\def\corrEmail{islee@ibs.re.kr}

\begin{document}
\twocolumn
\firstpage{1}

\title{Performance of an ultra-pure NaI(Tl) detector produced by an indigenously-developed purification method and crystal growth for the COSINE-200 experiment }
\author[]{\Authors} 
\address{} 
\correspondance{} 
\extraAuth{}
\maketitle

\begin{abstract}
		The COSINE-100 experiment has been operating with 106 kg of low-background NaI(Tl) detectors to test the results from the DAMA/LIBRA experiment, which claims to have observed dark matter. However, since the background of the NaI(Tl) crystals used in the COSINE-100 experiment is 2--3 times higher than that in the DAMA detectors, no conclusion regarding the claimed observation from the DAMA/LIBRA experiment could be reached. 
Therefore, we plan to upgrade the current COSINE-100 experiment to the next phase, COSINE-200, by using ultra-low background NaI(Tl) detectors. 
The basic principle  was already proved with the commercially available Astro-grade NaI powder from Sigma-Aldrich company. 
However, we have developed a mass production process of ultra-pure NaI powder at the Center for Underground Physics (CUP) of the Institute for Basic Science (IBS), Korea, using the direct purification of the raw NaI powder. We plan to produce more than 1,000\,kg of ultra-pure powder for the COSINE-200 experiment. 
With our crystal grower installed at CUP, we have successfully grown a low-background crystal using our purification technique for the NaI powder. We have assembled a low-background NaI(Tl) detector. In this article, we report the performance of this ultra-pure NaI(Tl) crystal detector produced at IBS, Korea. 
\tiny
 \keyFont{ \section{Keywords:} NaI(Tl) crystal; Dark matter; COSINE-200; Low-background detector; Purification} 

\end{abstract}



\section{Introduction}
Numerous astronomical observations support the theory that most of the matter in universe is the invisible dark matter, although an understanding of its nature and interactions remains elusive~\cite{Clowe:2006eq,Aghanim:2018eyx,Bertone:2016nfn}. 
Even though tremendous efforts have been made to search for dark matter, no definitive signals have been observed~\cite{Undagoitia:2015gya,Schumann:2019eaa}. The only exception is the DAMA experiment, which has observed an annual modulation of event rates using an array of NaI(Tl) detectors~\cite{Bernabei:2013xsa,Bernabei:2018jrt}. This observation could be interpreted as dark matter-nuclei interactions~\cite{Savage:2008er,Ko:2019enb}. 
However, this result has been the subject of a continuing debate because no other experimental searches have observed similar signals~\cite{Schumann:2019eaa,Workman:2022ynf}. 

Several experimental efforts using the same NaI(Tl) target materials are
currently underway~\cite{Kim:2014toa,sabre,Adhikari:2017esn,Fushimi:2018qzk,Coarasa:2018qzs,Amare:2018sxx,Suerfu:2019snq}. 
The COSINE-100 experiment is one such effort presently operating at the Yangyang underground laboratory in Korea, which has provided several exciting physics results~\cite{Ko:2019enb,Adhikari:2018ljm,COSINE-100:2021xqn,COSINE-100:2021poy,COSINE-100:2022dvc}. 
However, due to the approximately 2--3\ times higher background level, an unambiguous conclusion regarding the observation in the DAMA experiment using the same annual modulation signal 
has not been observed yet~\cite{COSINE-100:2019lgn,COSINE-100:2021zqh}.

As an effort to upgrade the ongoing COSINE-100 experiment for the next-phase COSINE-200 experiment, we have conducted an R\&D program aimed at producing a low-background NaI(Tl) detector to conclude on the observed signals from DAMA/LIBRA unambiguously. 
It includes the chemical purification of the raw NaI powder~\cite{Shin:2018ioq,Shin:2020bdq}, its crystal growth~\cite{Ra:2018kkl}, and detector assembly~\cite{Choi:2020qcj}. 
We have already proved the principle of a low-background NaI(Tl) detector using the commercially available low-background Astro-grade NaI powder from Sigma-Aldrich~\cite{Park:2020fsq}. 
As a next step, we have grown an NaI(Tl) crystal using our own NaO power produced using the mass purification process at IBS, Korea~\cite{Shin:2022}. 
This article reports the characteristics and performance of this indigenously-produced NaI(Tl) crystal.

\section{NaI purification and Crystal Growth}
The COSINE-200 detector requires extremely low levels of radioactive contamination in the materials used in the detector production. 
The major contributors to the background are the decays of $^{40}$K and $^{210}$Pb in the bulk NaI(Tl) crystal~\cite{cosinebg, Adhikari:2017gbj}. Because of the similarity in its chemical properties to those of Na, which is in the same periodic table group, K is the primary impurity contaminant, and its selective extraction from NaI powder is challenging.
We found that the fractional recrystallization method effectively reduces the K and Pb impurities~\cite{Shin:2018ioq}. 
In addition, using this method, the Ba concentration was significantly reduced, indicating a reduction of Ra impurities~\cite{Shin:2018ioq}.
Thus,we constructed a mass production facility at IBS, Daejeon, Korea, for producing ultra-pure NaI powder using the fractional recrystallization method on-site~\cite{Shin:2020bdq}. 
The facility has been operated with a maximum production rate of 35\,kg of ultra-pure powder in a single processing cycle of two weeks~\cite{Shin:2022}.
Using our purification facility, we have performed mass purification of the fractional recrystallization process using raw NaI powder from Merck (99.99(5)$\%$ purity). 
In this mass purification process of the NaI power, we have achieved a concentration of K of 6.4\,ppb and that of Pb below 0.3\,ppb~\cite{Shin:2022}, which are consistent with contamination levels of the Astro-grade powder.

The ultra-pure crystal was grown using a small-volume Kyropouls grower~\cite{Ra_2020}, which is the same grower used for growing the proof of principle low-background NaI(Tl) crystals using the commercial Astro-grade powder~\cite{Park:2020fsq}. 
In growing the crystal, 1.7\,kg of the purified NaI powder was loaded in a 12\,cm diameter, 10\,cm high quartz crucible. An NaI(Tl) crystal ingot, as shown in Figure~\ref{fig:crystal}(a), of $\sim$70\,mm diameter and $\sim$80\,mm high and having a 1.1\,kg mass, was grown in $\sim$24\,h . 
During the crystal growth, N$_2$ gas was continuously flushed using a thallium trap with a flow rate of 10\,L/m.

\begin{subfigure*}
  \centering
  \setcounter{subfigure}{0}
  \begin{minipage}[b]{0.4\textwidth}
    \includegraphics[width=.65\textwidth]{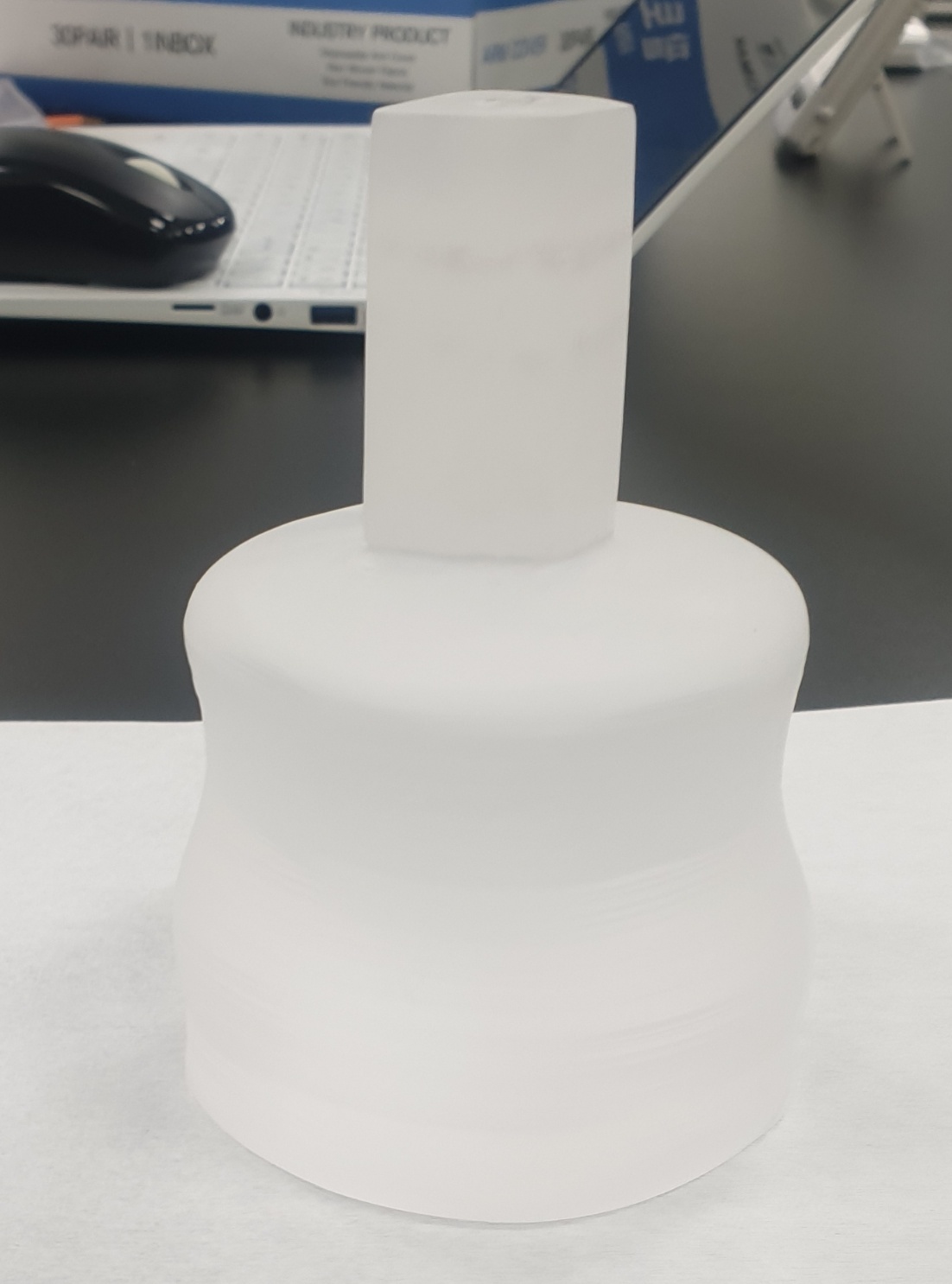}
    \caption{NaI(Tl) Crystal ingot}
        \label{fig:EncapDesign}
    \end{minipage}
    \setcounter{subfigure}{1}
 \begin{minipage}[b]{0.4\textwidth}
   \includegraphics[width=\linewidth]{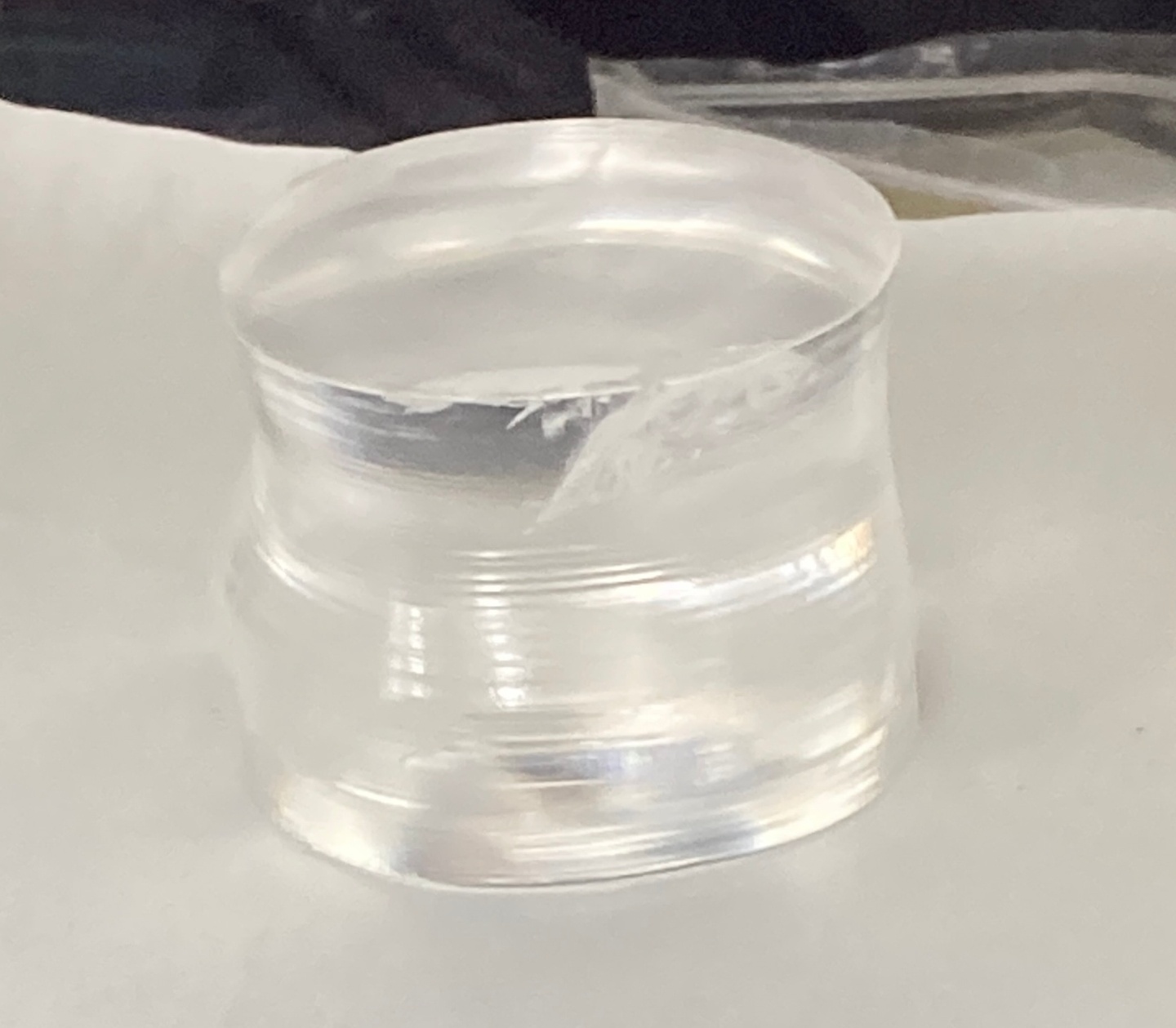}
   \caption{Cut and polished NaI(Tl) crystal }
        \label{fig:Encap}
    \end{minipage}   
   \setcounter{subfigure}{-1}
   \caption{Bare NaI(Tl) (NaI-037) crystal}
    \label{fig:crystal}
\end{subfigure*}

\section{Experimental setup}
\subsection{NaI(Tl) crystal}
The growth of the NaI(Tl) crystal (named NaI-037) was completed on January 18, 2021, using NaI powder purified IBS~\cite{Shin:2018ioq,Ra_2020}. The top and bottom sections of the crystal ingot were cut using a diamond bandsaw, as shown in Figure~\ref{fig:crystal}(b).
After cutting the top and bottom, the NaI-037 crystal is 70 mm in diameter and 51 mm in height.  
The flat top and bottom surfaces and a barrel-shaped side surface were polished using aluminum oxide films ranging from 400 to 8000 grit.
After polishing, the barrel was wrapped with a polytetrafluoroethylene (PTFE) film in several layers as a diffusive reflector.
A 3 mm thick copper casing with quartz windows at each end was encapsulated the crystal hermetically. 
Hamamatsu 3 inch photomultiplier tubes (PMTs), selected for high quantum efficiency (R12669SEL), were coupled via an optical interface to each end of the crystal. 
The entire assembly was performed in a glovebox, where the humidity was maintained to be less than 10 ppm (H$_{2}$O) using Ar gas and a molecular sieve trap.
Before the assembly, all parts were cleaned using diluted Citranox liquid with sonication and baked in a vacuum oven for more than 12 h. 
After assembly, the detector was delivered to the Yangyang underground laboratory (Y2L), which has $\sim$700\,m of rock overburden~\cite{COSINE-100:2017dsl}. From the crystal growth to Y2L delivery, it took less than three weeks and minimized the cosmogenic activation in the crystal. 

\subsection{Shielding structure}
\begin{figure}[!htb]
\begin{center}
    \includegraphics[width=0.5\textwidth]{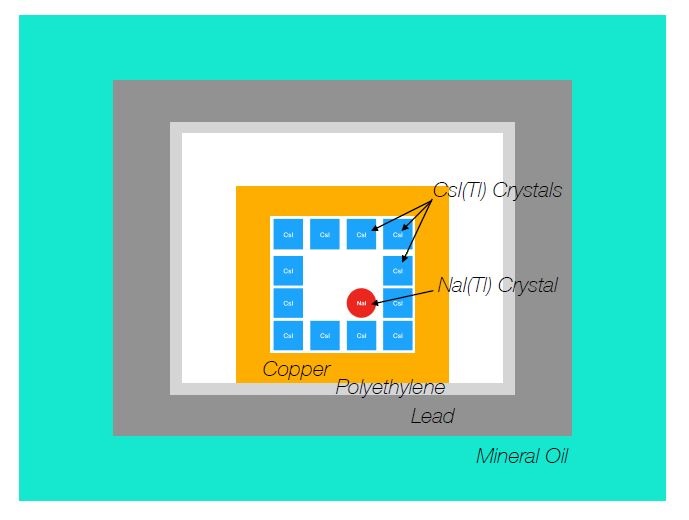}
\caption{A schematic view of the Y2L setup. The NaI-037 crystal (red circle) was installed inside the CsI(Tl) crystal array (blue squares).}
\label{fig:A6setup}
\end{center}
\end{figure}

The background contamination levels of the NaI-037 crystal were evaluated using, the same experimental apparatus as that used for the NaI(Tl) crystal R\&D at the Y2L~\cite{Park:2020fsq, adhikari16}.
It includes an array of 12 CsI(Tl) crystals surrounded by 10\,cm copper, 5\,cm polyethylene, 15\,cm lead, and 30\,cm liquid scintillator-loaded mineral oil~\cite{Lee:2005qr, LEE2007644} as a radiation shield.
The detector was installed inside the CsI(Tl) array, as shown in Figure~\ref{fig:A6setup}.

\subsection{Electronics}
The PMTs attached to the NaI-037 crystal had two readouts each, a high-gain signal from the anode and a low-gain signal from the fifth-stage dynode. The anode signal was amplified by a factor of 30, whereas the dynode signal was amplified by a factor of 100 using a custom-made preamplifier. 
The amplified signals were digitized by 500\,MHz, 12-bit flash analog-to-digital converters (FADCs).  
Triggers from the individual channels were generated by the field-programmable gate arrays (FPGAs) embedded in the FADC. 
The final trigger was generated in the trigger and clock board (TCB) when an anode signal corresponding to one or more photoelectrons (PEs) occurred in each PMT within a 200\,ns time window. 
The anode and dynode signals were recorded whenever the anode signal produced a trigger. 

Signals from the CsI(Tl) crystals were amplified by a factor of 10 and digitized in a charge-sensitive 62.5\,MHz FADC (SADC). 
The SADC provided the integrated charge and the time of the signal. An integration time of 2048 ns was used to record the CsI(Tl) signals considering their decay time.
The SADC channels did not generate triggers.

If the trigger condition was satisfied, the TCB sent trigger signals to the FADC and SADC to store the signals from the NaI(Tl) and the CsI(Tl) crystals.
The FADC stored an 8 $\mu$s long waveform starting approximately 2.4 $\mu$s before the time of the trigger position. 
The SADC stored the maximum integrated charge within an 8 $\mu$s search window. 
This system is similar to the one used in the COSINE-100 data acquisition~\cite{COSINE-100:2018rxe}. 

\begin{figure}[!htb]
\begin{center}
  \includegraphics[width=0.5\textwidth]{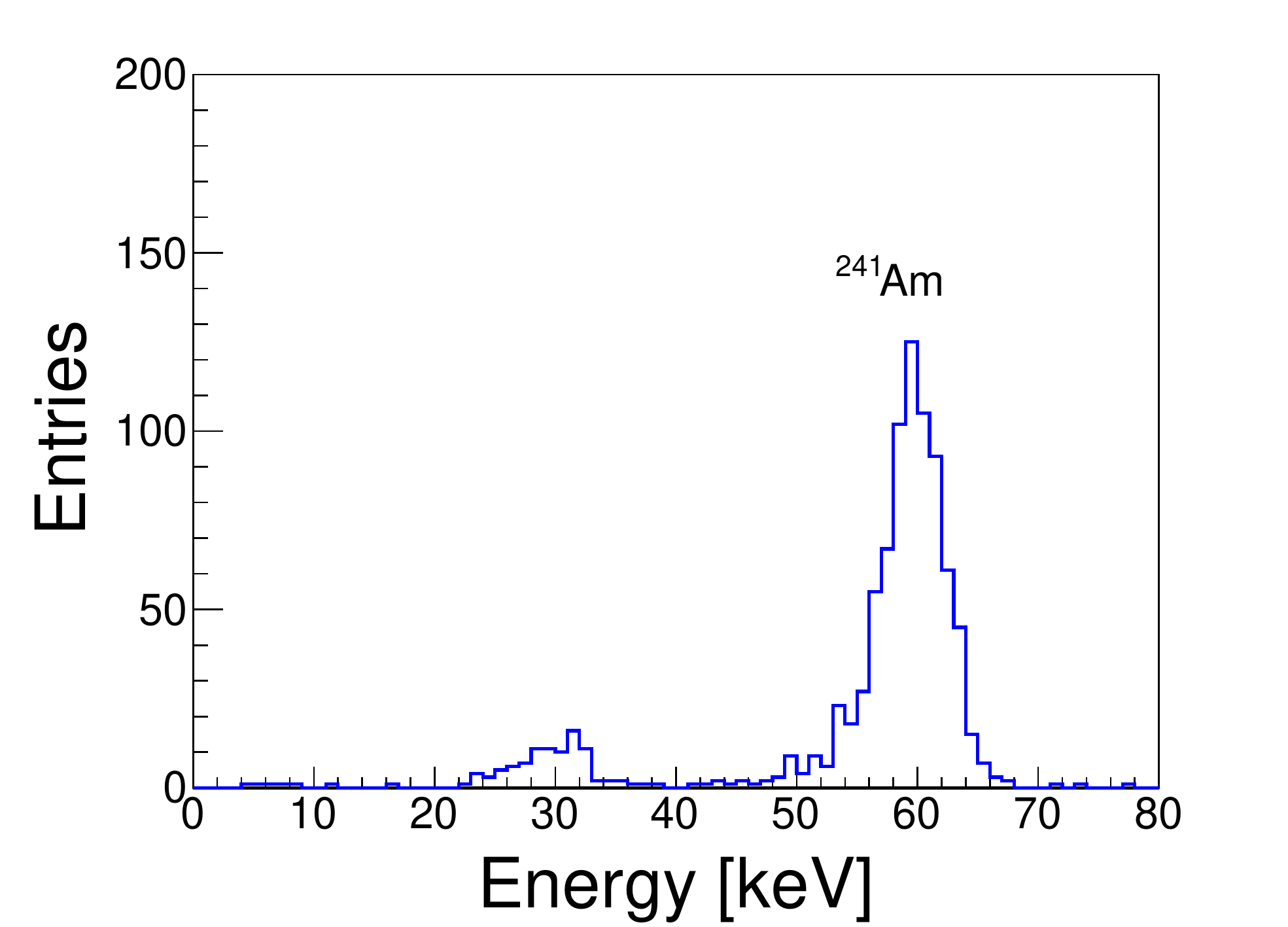}
 \caption{Anode energy distribution obtained using a $^{241}$Am source.}
\label{fig:Am241}
\end{center}
\end{figure}

\subsection{Energy calibration and light yields}
The energy calibration of the anode signal was done using a 59.54\,keV X-ray emitted from $^{241}$Am.
Figure~\ref{fig:Am241} shows the anode energy spectrum. A clear peak at 59.54\,keV resulting from the $^{241}$Am source is shown together with the $^{127}$I X-ray escape peak around 30\,keV. 
The dynode signal was calibrated using the photopeaks corresponding to $^{214}$Bi(609 keV) and  $^{40}$K(1460 keV) contaminants in the crystal.

The charge distribution of the single photoelectron (SPE) was obtained by identifying the isolated clusters at the decay tail of the 59.54 keV X-ray signal from the $^{241}$Am source (3-5 $\mu$s after the signal start). 
The light yield was determined from the ratio of the total deposited charge and the mean of the SPE charge for the 59.54 keV X-ray data. In this crystal, a light yield of 17.8$\pm$0.6 number of photoelectron (NPE)/keV was obtained. It is similar to the result for the NaI-036 crystal, which has the highest light yield among the previously developed low-background NaI(Tl) crystals using the Astro-grade powder~\cite{Park:2020fsq}. This light yield is also larger than those of the detectors used in the COSINE-100 and DAMA/LIBRA experiments, as summarized in table~\ref{table:MeasuredResult}.


\section{Understanding the background in the spectrum}

\subsection{$^{40}$K background}
$^{40}$K is one of the most problematic background sources in the search for weakly interacting massive particles (WIMP) using NaI(Tl) crystals.
The X-rays/Auger electrons from $^{40}$K decays produce 3.2\,keV energy signals, similar to the energy signals expected for a WIMP-nuclei interaction~\cite{cosinebg, Adhikari:2017gbj, amare:19may}.
The $^{40}$K decays also emit 1460\,keV $\gamma$ rays, which can escape from the NaI(Tl) crystal and hit the surrounding CsI(Tl) crystals, leading to a double coincidence with the 3.2\,keV X-rays.

\begin{figure}[!htb]
\begin{center}
    \includegraphics[width=0.5\textwidth]{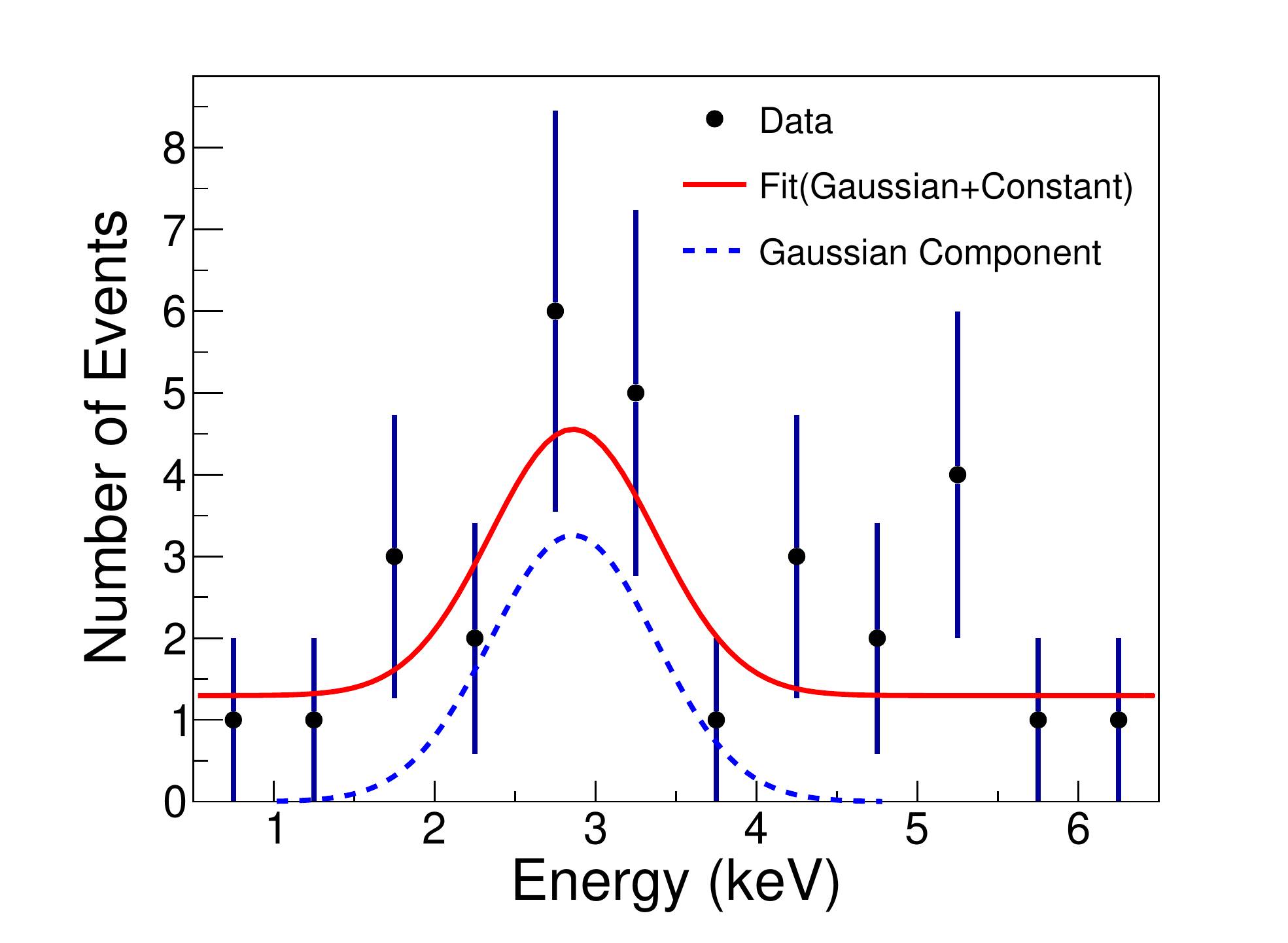}
    \caption{Energy deposition of the 3.2 keV $^{40}$K coincidence events in the NaI-037 crystal. The model of the energy spectrum assumes a combination of a Gaussian $^{40}$K signal and a constant background.}
\label{fig:K40fit}
\end{center}
\end{figure}

Figure~\ref{fig:K40fit} shows the tagged low-energy spectra from the NaI(Tl) crystal by requiring the detection of the 1460\,keV $\gamma$ ray in the CsI(Tl) crystals. 
The $^{40}$K background level in the NaI(Tl) crystal was determined by comparing the measured coincidence rate from a GEANT4-based simulated data, as described in Ref.~\cite{kwkim15}.
By accumulating more than six months of data, the K level was measured to be 8.3$\pm$4.6 ppb, which was compared with the other NaI(Tl) crystals listed in Table~\ref{table:MeasuredResult}.
It is well below our goal of 20\,ppb, consistent with the results from the DAMA/LIBRA crystals ~\cite{adhikari16, Bernabei:2008yh} and previously developed NaI-035 and NaI-036 crystals with the Astro-grade powder. 

\begin{figure}[!htb]
\begin{center}
    \includegraphics[width=0.5\textwidth]{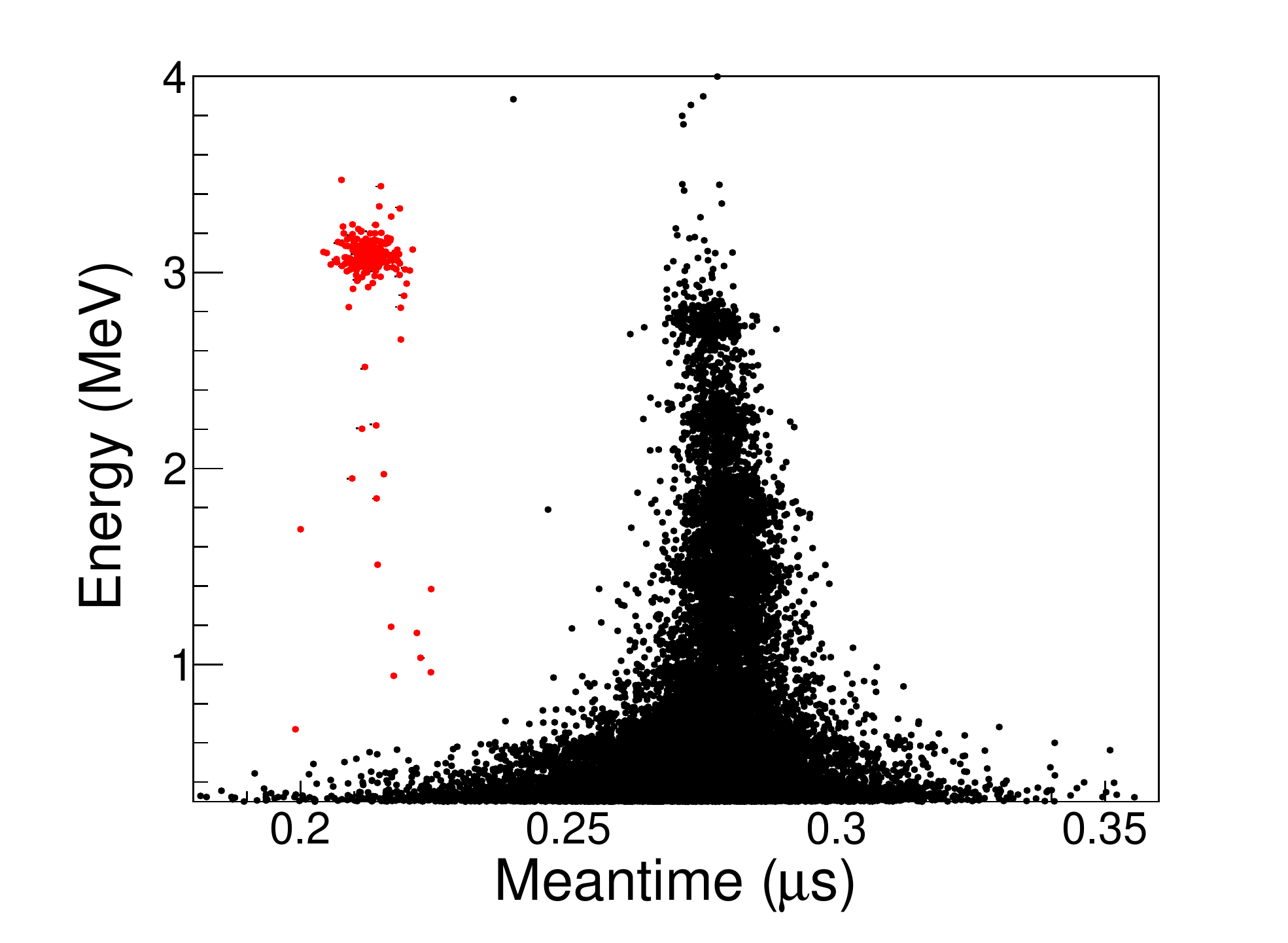}
    \caption{Scatter plot of the meantime versus the energy distribution events measured over 7.8 days for the NaI-037 crystal. The $\alpha$ events (red dots) and the $\gamma$/$\beta$ events (black dots) are separated clearly.}
\label{fig:AlphaScatter}
\end{center}
\end{figure}

\subsection{$\alpha$ analysis}
Alpha-induced events inside the NaI(Tl) crystals can be identified using the fast decay times of their corresponding signals. 
The charge-weighted duration time, called the meantime, is defined as
%
\begin{eqnarray}
\langle t \rangle = \frac{\Sigma_{i} A_{i} t_{i}}{\Sigma_{i} A_{i}},
 \label{eq:NuclearRecoil}
\end{eqnarray}
where A and t are the charge and time of the $\textit{i}$-th digitized bin of a signal waveform, respectively.
The meantime is estimated within 1500\,ns from the pulse starting timing.
Figure~\ref{fig:AlphaScatter} shows a scatter plot of the energy versus the meantime for the NaI-037 crystal.
In the figure, the populations of $\gamma$/$\beta$ and $\alpha$ events can be separated clearly due to the faster decay times of the $\alpha$-induced events.

\subsection{$^{210}$Pb background}
\begin{figure}[!htb]
\begin{center}
    \includegraphics[width=0.5\textwidth]{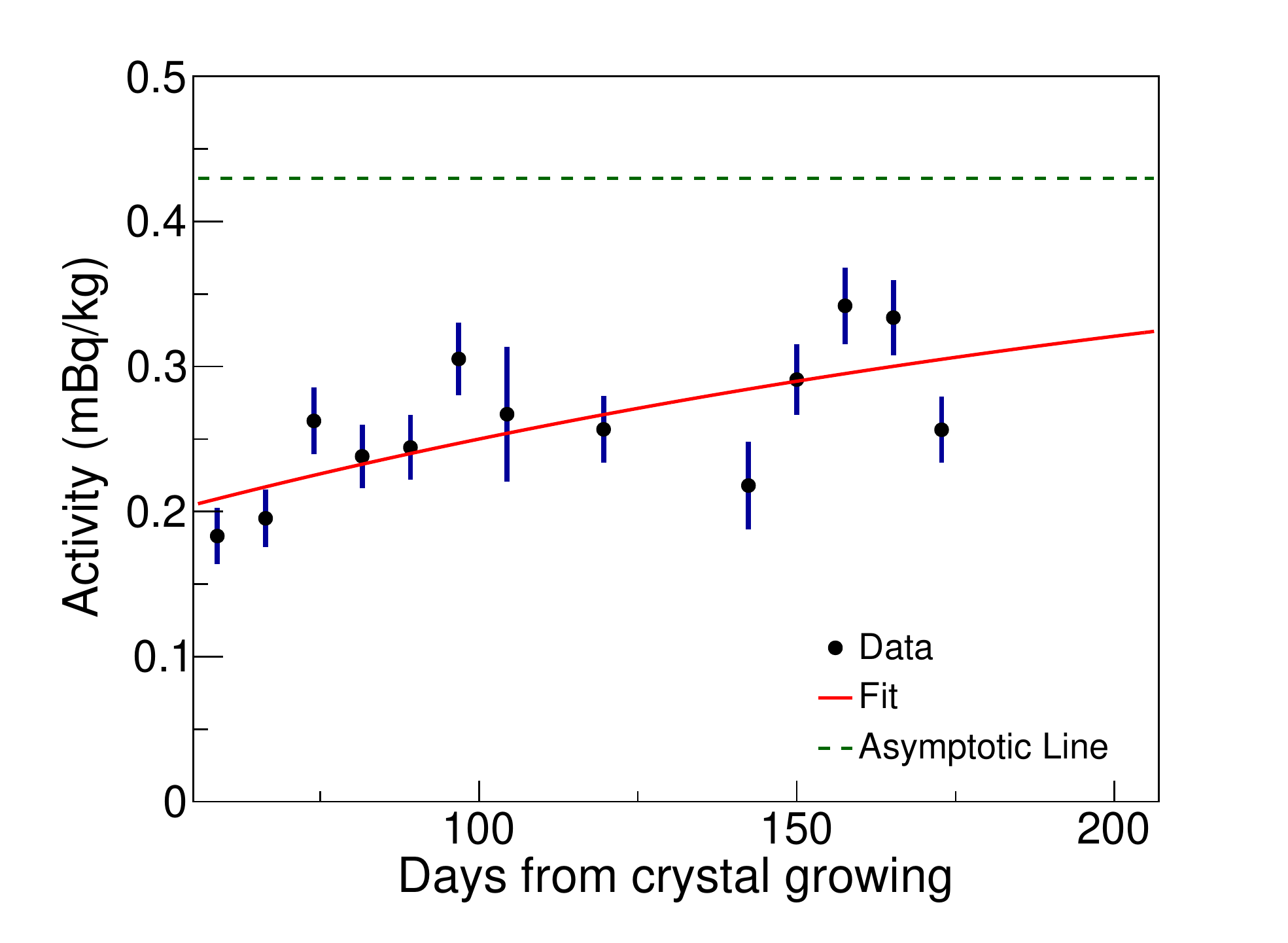}
    \caption{The total alpha rate in the NaI-037 crystal as a function of time,  modeled with $^{210}$Po assuming contamination of $^{222}$Rn (and/or $^{210}$Pb). The asymptotic line corresponds to the rate of total alpha events in the equilibrium state.}
\label{fig:TotAlp}
\end{center}
\end{figure}

In the NaI(Tl) crystal experiments, the dominant background source in the low-energy signal region is from $^{210}$Pb~\cite{Adhikari:2017gbj, Adhikari:2021rdm, Fushimi:2021mez}.
The $^{210}$Pb activity can be estimated from the alpha-decay studies, because the $\alpha$ decays of $^{210}$Po originate from the $\beta$ decays of $^{210}$Pb.
Due to the decay time of 200\,days of $^{210}$Po, the amount of $^{210}$Pb produced can be estimated using a time-dependent fit of the alpha rate as follows: 
\begin{eqnarray}
N = N_{Pb210}\left(1-e^{-(t-t_{0})/\tau_{Po210}}\right) + C,
 \label{eq:DecayEqn}
\end{eqnarray}
where $N$ is the total alpha rate, $N_{Pb210}$ is the amount of $^{210}$Pb at the equilibrium state, $t_{0}$ is the time difference between $^{210}$Pb contamination and the start time of data taking, $\tau_{Po210}$ is the mean lifetime of $^{210}$Po, approximately 200 days, and $C$ represents the long-lived components from $^{238}$U and $^{232}$Th chains.
Figure ~\ref{fig:TotAlp} shows the measured total alpha rates in the NaI-037 crystal over the detector running time. The $^{210}$Pb activity in the crystal was estimated to be 0.38$\pm$0.10\,mBq/kg, which is lower than the COSINE-100 crystals and is consistent with the activity in the NaI-036 crystal produced using the Astro-grade powder. However, this activity is slightly higher than the DAMA crystals and another crystal NaI-036 grown with the same Astro-grade powder. 
The 0.4\,mBq/kg level contamination of $^{210}$Pb is enough to reach 1 count/kg/keV/day background level, similar to the activity in the DAMA/LIBRA detectors, as described in Ref.~\cite{Park:2020fsq}.

\subsection{$^{232}$Th background}
Contaminants from the $^{228}$Th subchain in the $^{232}$Th family can be estimated by deploying the time-delayed $\alpha$--$\alpha$ coincident events of $^{220}$Rn and $^{216}$Po.
The alpha decay of $^{216}$Po has a half-life of 0.145 s following its production via alpha decay of $^{220}$Rn.
Owing to the short half-life of $^{216}$Po, it is straightforward to select two successive $\alpha$ particles with almost no random coincident events. 

The presence of the coincident events is shown in figure~\ref{fig:HeavyContaminant}(a) as the distribution of the time gap between those two $\alpha$ events.
The exponential component indicates the contamination from $^{232}$Th, corresponding to below 0.39 ppt (90$\%$ confidence level).
The $^{232}$Th concentration in the NaI-037 crystal is the lowest among the other NaI(Tl) crystals, as summarized in table~\ref{table:MeasuredResult}.

\subsection{$^{238}$U background}
$^{238}$U is one of the common radioisotopes because of its long half-life .
The $^{238}$U content in the background can be studied using the time-delayed $\beta$--$\alpha$ coincident events, similar to the calculation of the $^{232}$Th background.
This method exploits the $\alpha$ decay of $^{214}$Po with a half-life of  164.3 $\mu$s, while $^{214}$Bi, the parent particle of $^{214}$Po, undergoes $\beta$ decay.
Due to the 50 $\mu$s dead time of the trigger system, the coincident events with delay times greater than 50 $\mu$s can be tagged. 
The results are shown in figure~\ref{fig:HeavyContaminant}.
The $^{238}$U activity of NaI-037 was 1.02$\pm$0.58 ppt, similar to that observed for the other NaI(Tl) crystals, as given in Table~\ref{table:MeasuredResult}.

\begin{subfigure*}
  \setcounter{subfigure}{0}
    \begin{minipage}[b]{0.5\textwidth}
    \includegraphics[width=\linewidth]{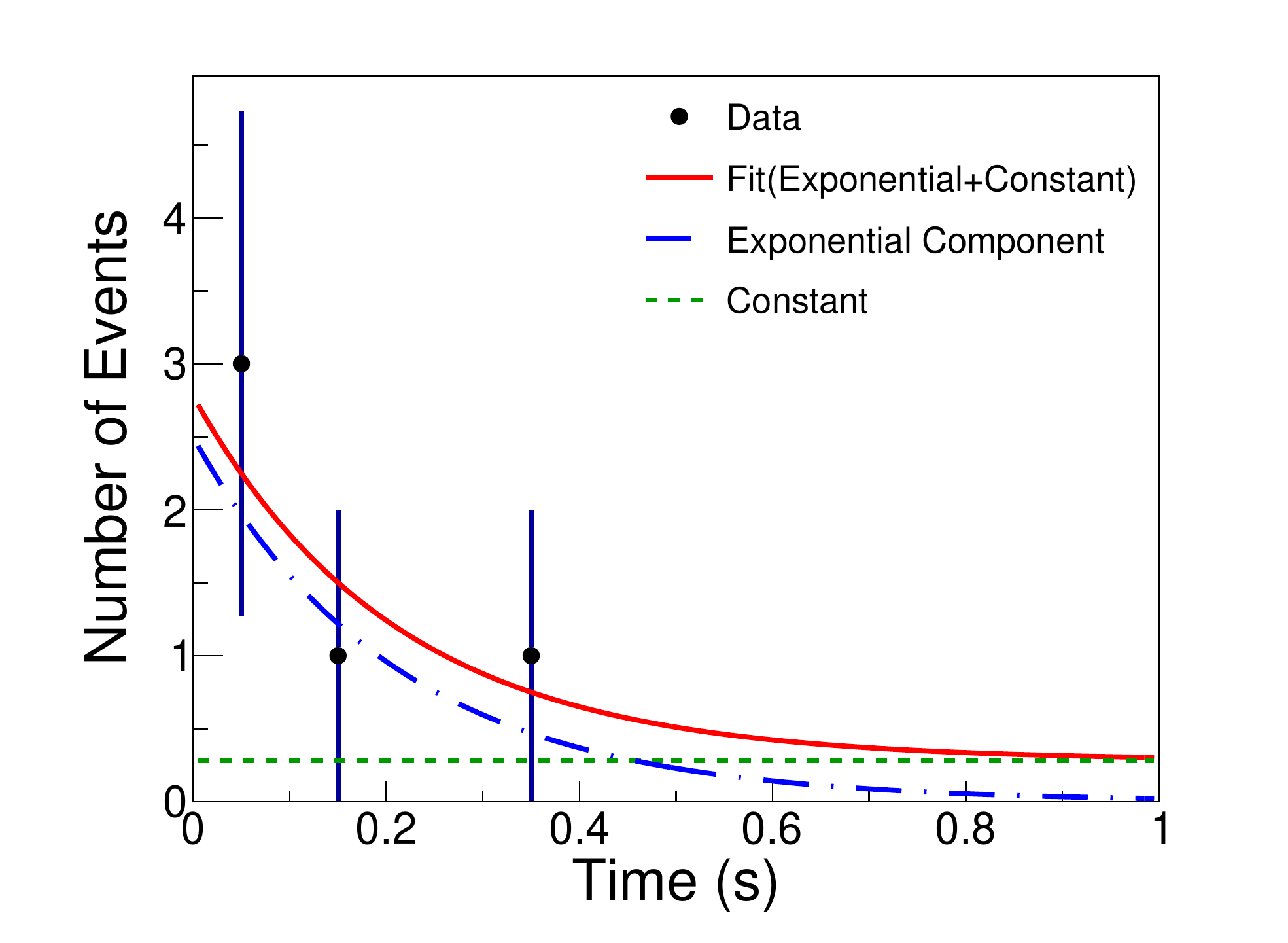}
    \caption{Time differen ofbetween two $\alpha$ decays of the $^{220}$Rn--$^{216}$Po decay chain.}
        \label{fig:HeavyContaminantTh228}
    \end{minipage}
    \setcounter{subfigure}{1}
    \hfill
 \begin{minipage}[b]{0.5\textwidth}
   \includegraphics[width=\linewidth]{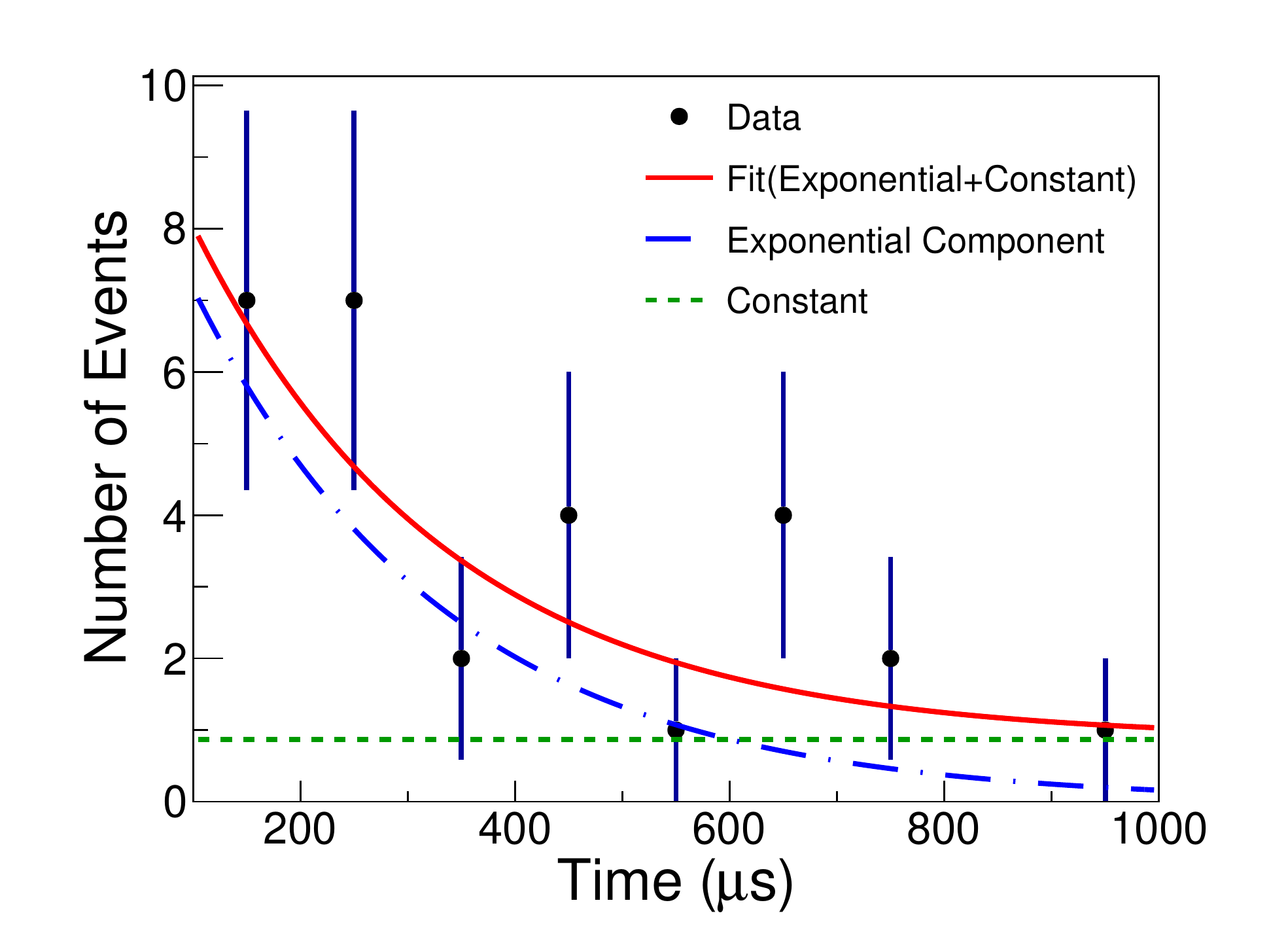}
   \caption{Time difference between the $^{214}$Po $\alpha$ decay and $^{214}$Bi $\beta$ decay.}
        \label{fig:HeavyContaminantU238}
    \end{minipage}   
   \setcounter{subfigure}{-1}
   \caption{Time difference distributions of data (black dots) and the exponential fits to them (red-solid line).}
    \label{fig:HeavyContaminant}
\end{subfigure*}

\begin{table*}[t!]
  \caption{Measured radioactive contaminants in the NaI-037 crystal, C6 of COSINE-100~\cite{cosinebg}, DAMA crystals~\cite{Bernabei:2008yh, bernabei12had}, and the previously grown NaI-035 and NaI-036 crystals using the Astro-grade powder~\cite{Park:2020fsq}.The upper limits are given at a 90$\%$ confidence level. }
  \begin{center}
    \renewcommand{\arraystretch}{1.5}
    \begin{tabular}{c c c c c c c}
      \hline \hline
      Crystal & Mass (kg) & LY (NPE/keV) & $^{40}$K (ppb) & $^{210}$Pb (mBq/kg) & $^{232}$Th (ppt) & $^{238}$U (ppt)\\ \Xhline{3\arrayrulewidth}
      NaI-037    & 0.71 & 17.8$\pm$0.6      & 8.3$\pm$4.6 & 0.44$\pm$0.09 & 0.2$\pm$0.3 & 1.0$\pm$0.6 \\ \hline
      NaI-035    & 0.61 & 11.8$\pm$1.8 & $<$42         & 0.01$\pm$0.02 & 1.7$\pm$0.5 & 0.9$\pm$0.3\\
      NaI-036    & 0.78 & 17.1$\pm$0.5 & $<$53         & 0.42$\pm$0.27 & $<$4.9 & 36.5$\pm$3.9\\
      COSINE-100 & 12.5 & 14.6$\pm$1.5 & 16.8$\pm$2.5  & 1.87$\pm$0.09 & 0.7$\pm$0.2 & $<$0.02\\
      DAMA       & 9.7  & 5--10        & $<$20         & 0.01--0.03 & 0.5--7.5 & 0.7--10\\
      \hline \hline
      
    \end{tabular}
  \end{center}
  
  \label{table:MeasuredResult}
\end{table*}
 
\subsection{External Background}
Because of the small size of the NaI-037 crystal and no liquid scintillator active veto, a significantly higher background contribution is expected from the external background compared to those found in the COSINE-100 crystals~\cite{cosinebg,Adhikari:2021rdm}.
The PMTs attached to the NaI(Tl) and the CsI(Tl) crystals are the primary sources of external background. 
In this study, the external background contributions were simulated using the GEANT4-based simulation toolkit used for the COSINE-100 background modeling~\cite{cosinebg, Adhikari:2021rdm}.

\subsection{Cosmogenic radionuclides}
The cosmogenic production of radioactive isotopes in
the NaI(Tl) crystal is mainly due to long-lived nuclides such as $^{3}$H and $^{22}$Na~\cite{cosinebg, BARBOSADESOUZA2020102390}.
The NaI-037 crystal was grown in Daejeon, Korea (70 m in altitude) and delivered underground within a month. 
Based on the previous study, one-month exposure time near sea level can produce 0.004 mBq/kg of $^{3}$H and 0.05 mBq/kg of $^{22}$Na~\cite{BARBOSADESOUZA2020102390}, respectively.

\begin{subfigure*}
    \setcounter{figure}{9}
  \setcounter{subfigure}{0}
    \begin{minipage}[b]{0.5\textwidth}
    \includegraphics[width=\linewidth]{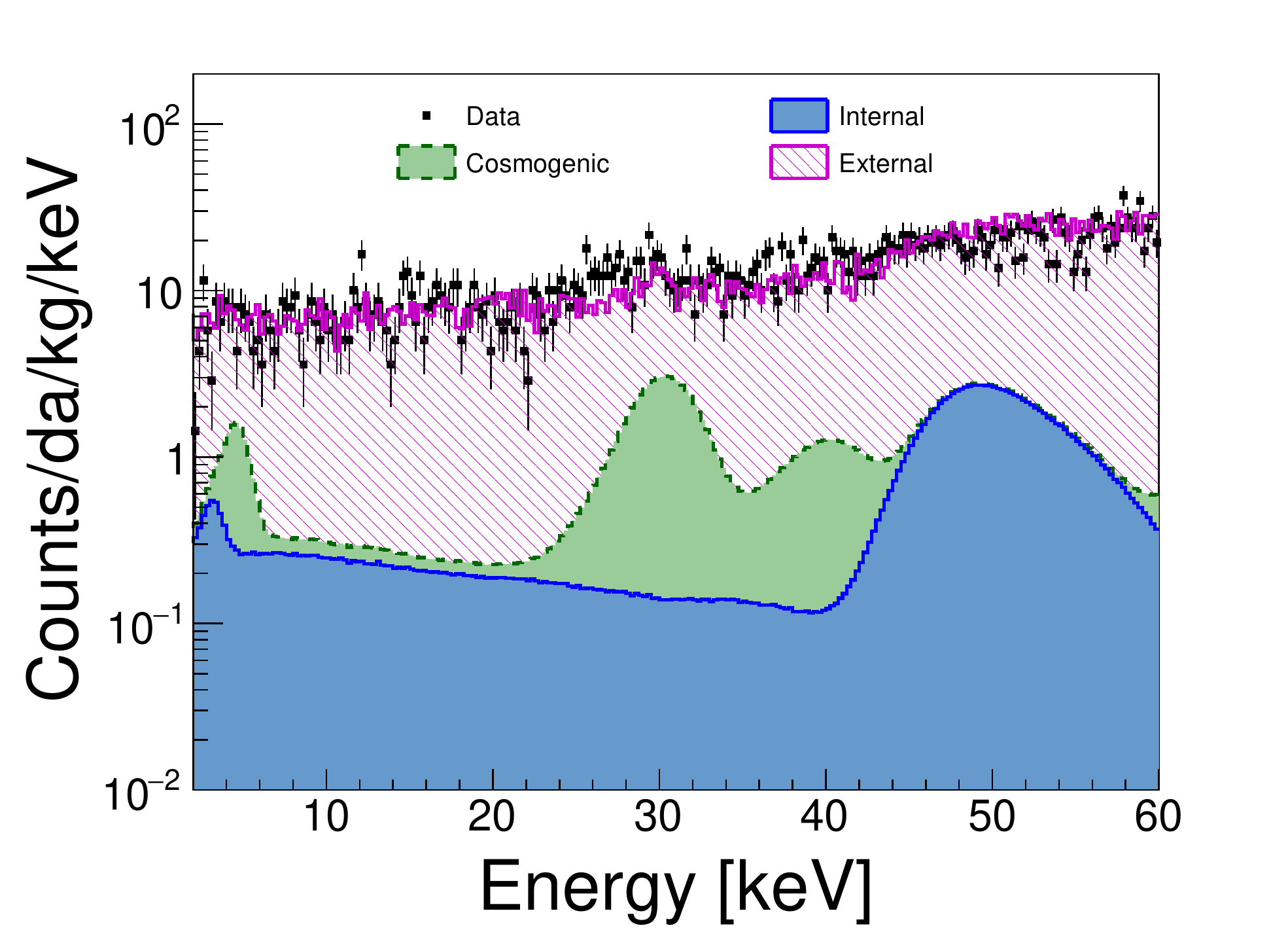}
        \caption{Single-hit low-energy (2--60 keV)}
        \label{fig:ModelingSinLow}
    \end{minipage}
     \setcounter{figure}{9}
    \setcounter{subfigure}{1}
    \hfill
 \begin{minipage}[b]{0.5\textwidth}
   \includegraphics[width=\linewidth]{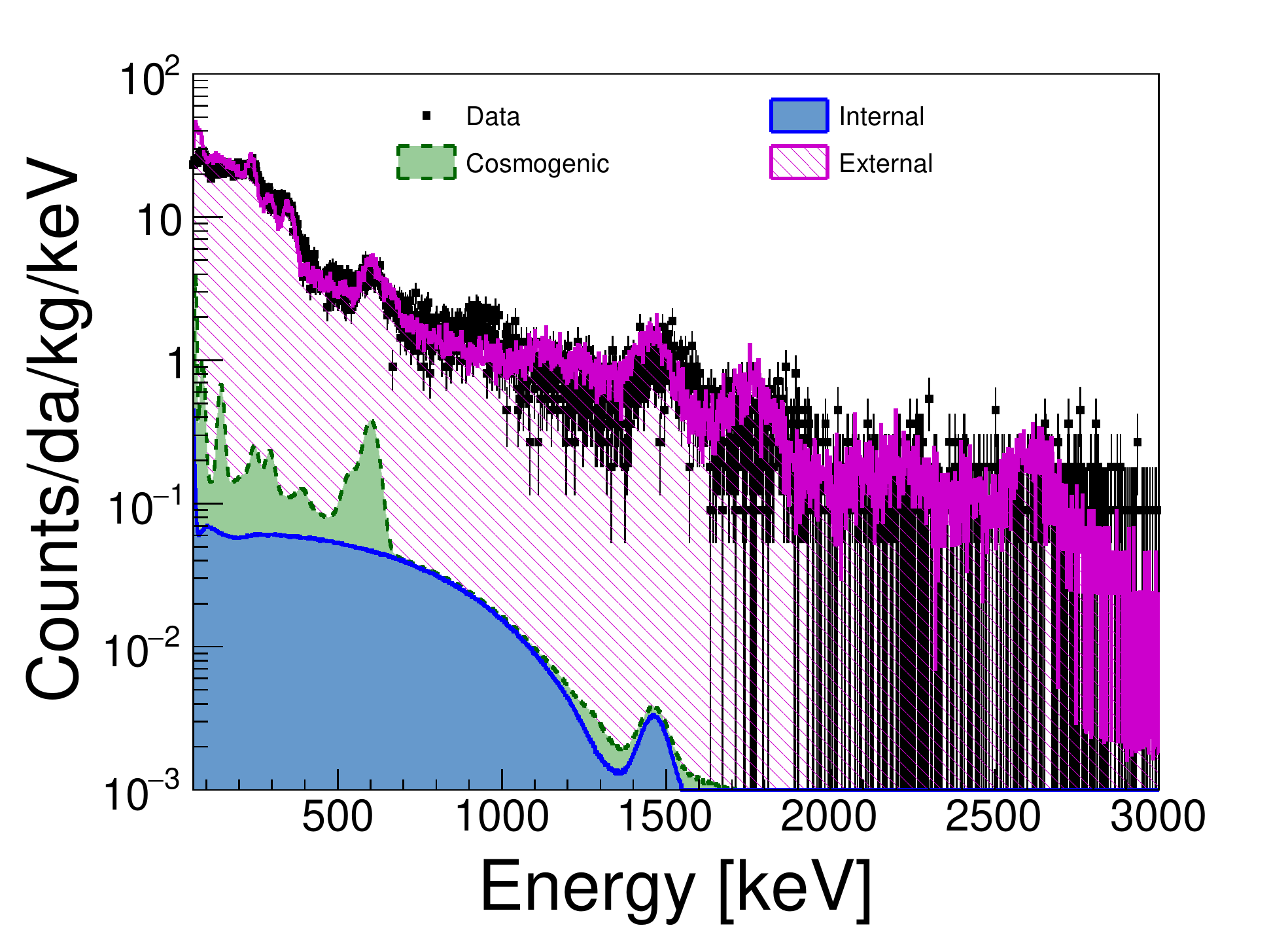}
     \caption{Single-hit high-energy (60--3000 keV)}
        \label{fig:ModelingSinHigh}
    \end{minipage}   
 \setcounter{figure}{9}
   \setcounter{subfigure}{2}
    \begin{minipage}[b]{0.5\textwidth}
    \includegraphics[width=\linewidth]{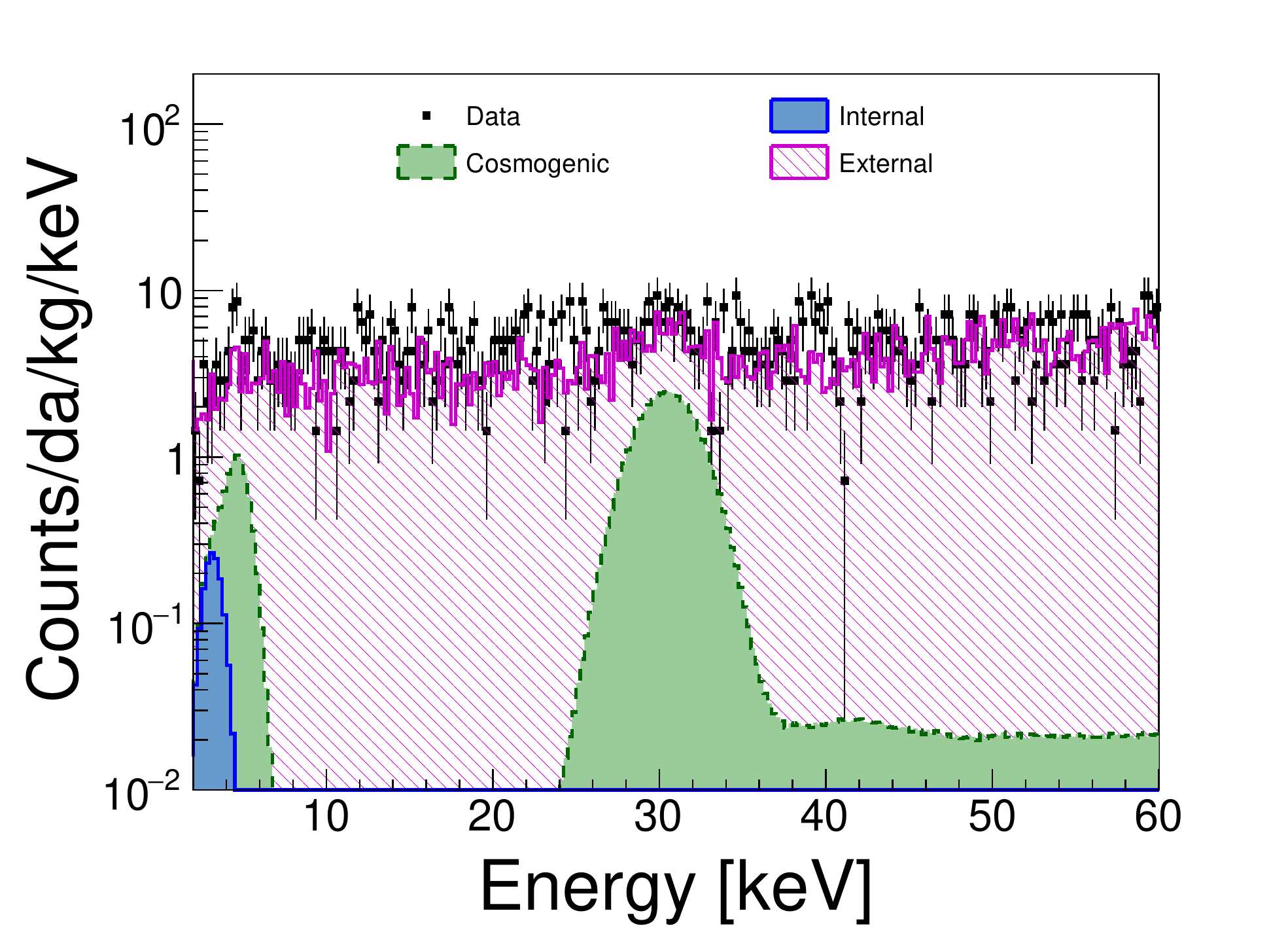}
        \caption{Multiple-hit low-energy (2--60 keV)}
        \label{fig:ModelingMulLow}
    \end{minipage}
    \setcounter{figure}{9}
    \setcounter{subfigure}{3}
    \hfill
    \begin{minipage}[b]{0.5\textwidth}
      \includegraphics[width=\linewidth]{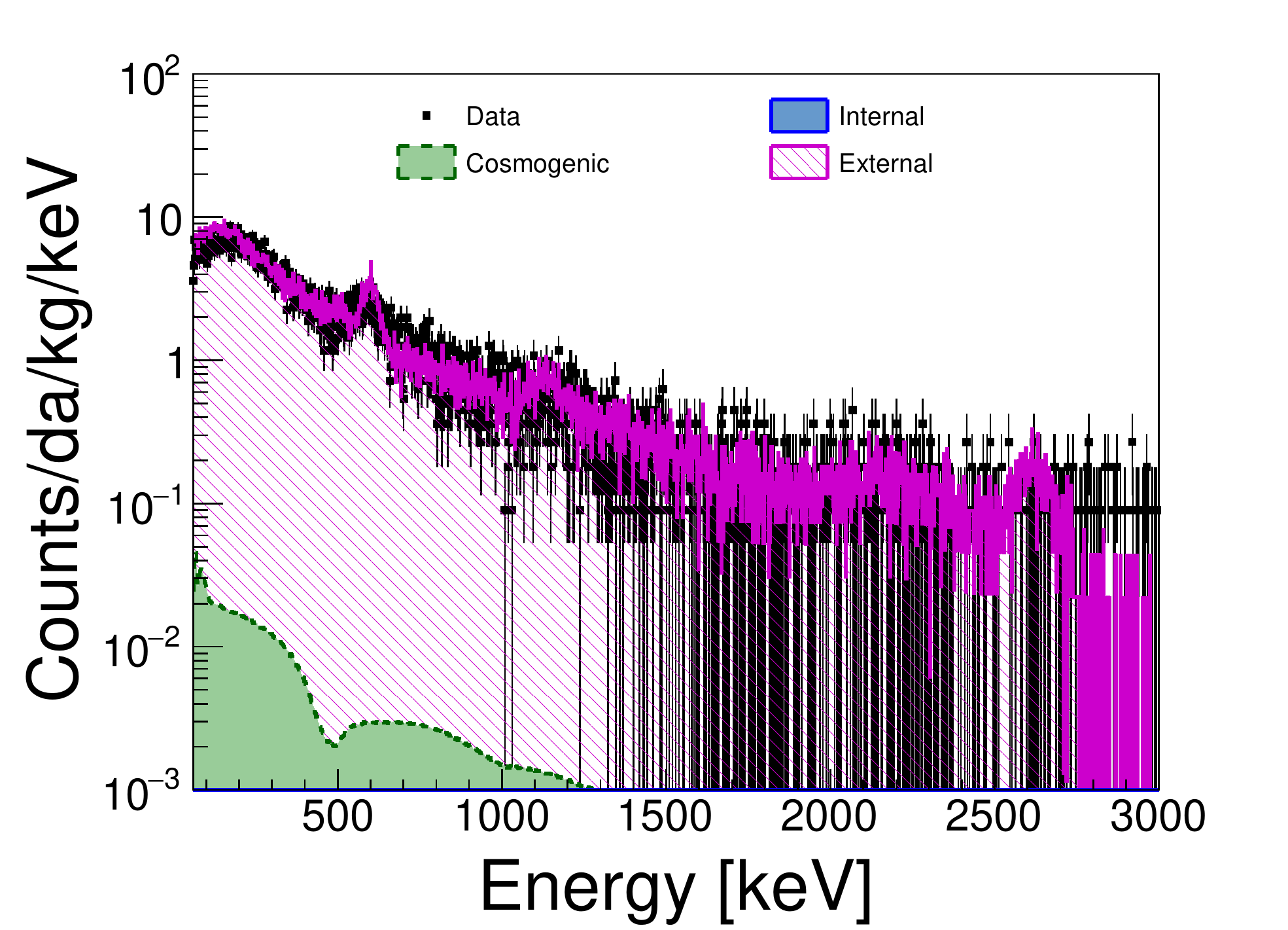}
        \caption{Multiple-hit high-energy (60--3000 keV)}
        \label{fig:ModelingMulHigh}
    \end{minipage}   
    \setcounter{figure}{9}
   \setcounter{subfigure}{-1}
    \caption{Measured single-hit and multiple-hit background spectra of the NaI-037 (black point) crystal fitted with the different simulated background components using a simultaneous fit of four channels using the log-likelihood method. The external component (purple-hatched area) is the dominant contributor.}
    \label{fig:modeling}
\end{subfigure*}

\section{Background modeling}
For a quantitative understanding of the background in the NaI-037 crystal, GEANT4-based simulation, developed for the background modeling of the COSINE-100 NaI(Tl) crystals~\cite{cosinebg, Adhikari:2021rdm} and also used in the previously grown crystals using the Astro-grade powder~\cite{Park:2020fsq}, was performed.
The input values of the contamination levels are obtained from Table~\ref{table:MeasuredResult}. 
A simultaneous fit was done to the single-hit low energy (3--60 keV), single-hit high energy (60 keV--3 MeV), multiple-hit low energy, and multiple-hit high energy events using the log-likelihood method. 
A multiple-hit event corresponds to one or more coincident hits in any of the surrounding CsI(Tl) crystals.
The backgrounds from the PMTs attached to the NaI(Tl) and CsI(Tl) crystals were measured using a high-purity germanium detector~\cite{cosinebg, Adhikari:2017gbj}.
These values were constrained to be within 50$\%$ of the measured result because the exact locations of such radioisotopes are uncertain. 
The long-lived cosmogenic radioisotopes were constrained to be within 50$\%$ of their calculation production values whereas the other short-lived cosmogenic components were floated.
Figure~\ref{fig:modeling} and Table~\ref{table:fitresult} show the fitted results for the NaI-037 crystal on all simulated background components and the summary of the fitted radioactive contaminants, respectively.
The overall energy spectra match the data for the single-hit and multiple-hit events satisfactorily. 
The level of the fitted internal components is similar to the previously grown NaI-036 crystal~\cite{Park:2020fsq}.

The expected background level in the COSINE-200 crystal can be studied from the simulated background by assuming a 12.5\,kg detector in the COSINE-100 shielding, as described in Ref.~\cite{Park:2020fsq}. If the measured backgrounds, given in Table~\ref{table:fitresult} for the simulated study, are considered, a background level of approximately 0.5\,counts/kg/keV/day in the 1--6\,keV energy region is obtained, which is similar to the result for the NaI-036 crystal in the previous study~\cite{Park:2020fsq}. This is a slightly higher background level than observed from the NaI-035 crystal owing to the higher $^{210}$Pb contamination. However, it is still less than 1 count/kg/keV/day, the target background level for the COSINE-200 experiment.

\begin{table}[t!]
  \caption{Summary of the fitted radioactive contaminants in the modeling of the NaI-037 crystal.}
  \begin{center}
    \begin{tabular}{c c c}
      \hline \hline
      Background source & Isotope & Activity (mBq/kg)\\\hline
      
      \multirow{4}{*}{Internal}  & $^{238}{\rm U}$  &$0.025\pm0.35$ \\
                                 & $^{228}{\rm Th}$  &$0.0065\pm0.00025$ \\
                                 & $^{40}{\rm K}$  & $0.17\pm0.047$\\
                                 & $^{210}{\rm Pb}$  & $0.36\pm0.11$\\ \hline

      \multirow{10}{*}{Cosmogenic} & $^{125}{\rm I}$  & $0.40\pm0.0015$\\
                                   & $^{121}{\rm Te}$  & $0.80\pm 0.0029$\\
                                   & $^{121m}{\rm Te}$  & $0.063\pm0.0096$\\
                                   & $^{123m}{\rm Te}$  & $0.045\pm0.099$\\ 
                                   & $^{125m}{\rm Te}$  & $0.14\pm0.011$\\
                                   & $^{127m}{\rm Te}$  & $0.16\pm 0.10$\\
                                   & $^{109}{\rm Cd}$  & $0.0071\pm0.0010$\\ 
                                   & $^{113}{\rm Sn}$  & $0.020\pm0.00094$\\
                                   & $^{22}{\rm Na}$  & $0.050\pm0.010$\\
                                   & $^{3}{\rm H}$  & $0.0037\pm0.0097$\\ \hline
      \multirow{3}{*}{NaI PMTs}  & $^{238}{\rm U}$  & $48.83\pm5.90$\\
                                 & $^{228}{\rm Th}$  & $23.80\pm5.70$\\
                                 & $^{40}{\rm K}$  & $58.07\pm17.82$\\\hline
      \multirow{3}{*}{CsI PMTs}  & $^{238}{\rm U}$  & $27.64\pm6.15$\\
                                 & $^{228}{\rm Th}$  & $24.18\pm6.10$\\
                                 & $^{40}{\rm K}$  & $378.28\pm17.74$\\\hline\hline

    \end{tabular}
  \end{center}
  
  \label{table:fitresult}
\end{table}

\section{Conclusion}
In this article, we presented the performance of the first ultra-low background NaI(Tl) crystal produced using the direct purification of the NaI powder in our facility as a part of a program for the next-generation COSINE-200 experiment. 
The results of this study show a similar quantity of internal background contamination in the crystals grown using commercial Astro-grade powder. 
It indicates that the direct powder purification and crystal growth procedures employed at our facility can provide suitable NaI(Tl) crystals for the COSINE-200 experiment.
Based on the experience of developing ultra-pure NaI(Tl) crystals, we are moving to full-size crystal growth with our purified powder.

\section*{Acknowledgments}
We thank Korea Hydro and Nuclear Power Co., Ltd. (KHNP) for providing the underground laboratory space at Yangyang. This work is supported by the Institute for Basic Science (IBS) under the project code IBS-R016-A1. 

\newpage

\bibliographystyle{Frontiers-Vancouver}
\bibliography{dm}

\end{document}